\documentclass[twocolumn,prb,showpacs]{revtex4}

\usepackage{graphicx}
\usepackage{epsfig}
\usepackage{dcolumn}
\usepackage{bm}
\usepackage{amsmath}

\begin{document}
\title{Hall Voltage with the Spin Hall Effect}
\author{Yuriy V. Pershin}
\email{pershin@physics.ucsd.edu}
\author{Massimiliano Di Ventra}
\email{diventra@physics.ucsd.edu} \affiliation{Department of
Physics, University of California, San Diego, La Jolla, California
92093-0319}

\begin{abstract}
The spin Hall effect does not generally result in a charge Hall
voltage. We predict that in systems with inhomogeneous electron
density in the direction perpendicular to main current flow, the
spin Hall effect is instead accompanied by a Hall voltage. Unlike
the ordinary Hall effect, we find that this Hall voltage is
quadratic in the longitudinal electric field for a wide range of
parameters accessible experimentally. We also predict spin
accumulation in the bulk and sharp peaks of spin-Hall induced charge
accumulation near the edges. Our results can be readily tested
experimentally, and would allow the electrical measurement of the
spin Hall effect in non-magnetic systems and without injection of
spin-polarized electrons.
\end{abstract}

\pacs{72.25.Dc, 71.70.Ej}

\maketitle

Currently, much attention is given to studies of the spin Hall
effect, which allows to polarize electron spins without magnetic
fields and/or magnetic materials
\cite{natureSHC,r0,r1,r2,r3,Rashba,Rashba1,r4,r5,r6,r7,r8,r9,r10,Kato,Wunderlich,Valenzuela}.
In the spin Hall effect, electrically induced electron spin
polarization accumulates near the edges of a channel and is zero
in its central region. This effect is caused by deflection of
carriers moving along an applied electric field by
extrinsic\cite{r1} and/or intrinsic\cite{r3} mechanisms. In a
non-magnetic homogeneous system, spin accumulation is {\em not}
accompanied by a charge Hall voltage because two spin Hall
currents cancel each other \cite{natureSHC}. The absence of Hall
voltage leads to difficulties in probing the spin Hall effect,
since measuring a charge accumulation is much easier than
measuring a spin accumulation. Recently, the spin Hall effect has
been observed both optically \cite{Kato,Wunderlich} and
electrically \cite{Valenzuela}. In the latter case, a charge
accumulation has been created through injection of spin-polarized
electrons into the sample \cite{Valenzuela}.

In the present Letter, we predict that in a system with an
inhomogeneous electron density profile in the direction
perpendicular to the direction of main current flow, the spin Hall
effect results in {\em both} spin and charge accumulations. The
pattern of charge accumulation is determined by the interplay of two
mechanisms. The first mechanism of charge accumulation is based on
the dependence of spin-up and spin-down currents on local spin-up
and spin-down densities. Spin currents, outgoing from regions with
higher densities, are not fully compensated by incoming currents,
therefore, a charge accumulation appears. This mechanism is
primarily responsible for the non-zero Hall voltage. The second
mechanism of charge accumulation is related to scattering of spin
currents on sample boundaries which act like obstacles. Like in the
case of Landauer resistivity dipoles \cite{Landauer}, this
scattering leads to formation of local charge accumulation, which is
also expected in traditionally-studied spin Hall systems. In
addition, we show that in systems with inhomogeneous electron
density the spin accumulation appears not only near the sample
boundaries, but also in the bulk. Our approach does not involve any
use of magnetic materials and fields, therefore, the spin Hall
effect can be measured electrically in completely non-magnetic
system and without injection of spin-polarized electrons.

\begin{figure}[b]
\includegraphics[angle=270,width=7.5cm]{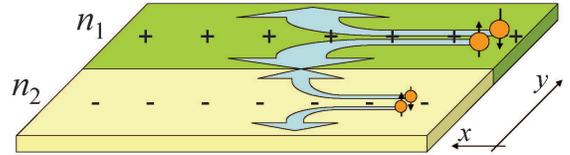}
\caption{\label{fig1} (Color online) Spin Hall effect in a system
with a step profile of the electron density in $y$ direction,
$n_1>n_2$. Spin currents through the boundary between $n_1$ and
$n_2$ do not cancel each other, resulting in a Hall voltage.}
\end{figure}

To illustrate this effect, let us begin by considering a system
having a step profile of electron density, as shown in Fig.
\ref{fig1}. There are several possible ways to fabricate such a
system including density depletion by an electrode, inhomogeneous
doping \cite{YP}, or variation of the sample height. What is
important for us is that the perpendicular (in $y$ direction) spin
currents are different in the regions with different electron
density. Then, if we consider currents passing through the boundary
separating regions with different charge densities ($n_1$ and
$n_2$), it is clear that the spin current from the region with
higher electron density has a larger magnitude than the current in
the reverse direction. The difference in currents implies charge
transfer through the boundary and formation of a dipole layer.

Let us now provide a quantitative analysis of this effect. We employ
a two-component drift-diffusion model \cite{flatte,PDV}, and in
order to find a self-consistent solution, we supplement the
drift-diffusion equations with the Poisson equation. In our
drift-diffusion calculation scheme, the inhomogeneous charge density
profile $n(y)$ is defined via an assigned positive background
density profile $N(y)$ (such as the one in Fig.~\ref{fig1}), which,
as discussed above, can be obtained in different ways. Assuming
homogeneous charge and current densities in $x$ direction and
homogeneous $x$-component of the electric field in both $x$ and $y$
directions, we can write a set of equations including only $y$ and
$t$ dependencies:

\begin{equation}
e\frac{\partial n_{\uparrow (\downarrow)}}{\partial
t}=\textnormal{div}  j_{y,\uparrow
(\downarrow)}+\frac{e}{2\tau_{sf}}\left(n_{\downarrow
(\uparrow)}-n_{\uparrow (\downarrow)} \right), \label{contEq}
\end{equation}
\begin{equation}
j_{y,\uparrow (\downarrow)}=\sigma_{\uparrow (\downarrow)}
E_y+eD\nabla  n_{\uparrow (\downarrow)}\pm \gamma I_{x,\uparrow
(\downarrow)} , \label{currentEq}
\end{equation}
and
\begin{equation}
\textnormal{div}E_y=\frac{e}{\varepsilon\varepsilon_0}\left(
N(y)-n\right), \label{puaeq}
\end{equation}
where $-e$ is the electron charge, $n_{\uparrow (\downarrow)}$ is
the density of spin-up (spin-down) electrons, $j_{y,\uparrow
(\downarrow)}$ is the current density, $\tau_{sf}$ is the spin
relaxation time, $\sigma_{\uparrow (\downarrow)}=en_{\uparrow
(\downarrow)}\mu$ is the spin-up (spin-down) conductivity, $\mu$ is
the mobility, $D$ is the diffusion coefficient, $\epsilon$ is the
permittivity of the bulk, and $\gamma$ is the parameter describing
deflection of spin-up (+) and spin-down (-) electrons. The current
$I_{x,\uparrow (\downarrow)}$ in $x$-direction is coupled to the
homogeneous electric field $E_0$ in the same direction as
$I_{x,\uparrow (\downarrow)}=en_{\uparrow (\downarrow)}\mu E_0$. The
last term in Eq. (\ref{currentEq}) is responsible for the spin Hall
effect.

Equation (\ref{contEq}) is the continuity relation that takes into
account spin relaxation, Eq. (\ref{currentEq}) is the expression for
the current in $y$ direction which includes drift, diffusion and
spin Hall effect components, and Eq. (\ref{puaeq}) is the Poisson
equation. It is assumed that $D$, $\mu$, $\tau_{sf}$ and $\gamma$
are equal for spin-up and spin-down electrons.~\cite{prec} In our
model, as it follows from Eq. (\ref{currentEq}), the spin Hall
correction to spin-up (spin-down) current (the last term in Eq.
\ref{currentEq}) is simply proportional to the local spin-up
(spin-down) density. All information about microscopic mechanisms
for the spin Hall effect is therefore lumped in the parameter
$\gamma$.

Combining equations (\ref{contEq}) and (\ref{currentEq}) for
different spin components we can get the following equations for
electron density $n=n_{\uparrow}+n_{\downarrow}$ and spin density
imbalance $P=n_{\uparrow}-n_{\downarrow}$:
\begin{equation} \frac{\partial
n}{\partial t}=\frac{\partial}{\partial y} \left[ \mu n E_y + D
\frac{\partial n}{\partial y} +\gamma P \mu E_0 \right] \label{CC}
\end{equation}
and
\begin{equation} \frac{\partial P}{\partial t}=\frac{\partial}{\partial y}
\left[ \mu P E_y + D \frac{\partial P}{\partial y} +\gamma n \mu
E_0 \right]-\frac{P}{\tau_{sf}}. \label{Peq}
\end{equation}

{\it Analytical solution --} Before solving Eqs.
(\ref{puaeq})-(\ref{Peq}) numerically, let us try to find analytical
solutions in specific cases. This will help us in the discussion of
the numerical results. An analytical steady-state solution of these
equations can indeed be found for the case of exponential density
profile in a system which is infinite in the $y$ direction.

The structure of Eqs. (\ref{puaeq})-(\ref{Peq}) allows us to
select a solution in the form
\begin{eqnarray}
n=N(y)=Ae^{\alpha y} \label{s1}, \\ P=Ce^{\alpha y},
\\E_y=\textnormal{const},\label{s2}
\end{eqnarray}
where $A$, $C$ and $\alpha$ are constants ($A$ and $\alpha$ are
assigned). This solution corresponds to constant spin polarization
$p=P/n$. Substituting Eqs. (\ref{s1})-(\ref{s2}) into Eqs.
(\ref{CC}) and (\ref{Peq}) (note that the Poisson equation
(\ref{puaeq}) is automatically satisfied) we obtain
\begin{eqnarray}
\mu E_y A+D\alpha A+\gamma \mu E_0C=0,\label{eq123}
\\ \mu E_y \alpha
C+D\alpha^2C+\gamma \mu E_0 \alpha A-\frac{C}{\tau_{sf}}=0.
\end{eqnarray}
From these equations, eliminating $E_y$, we find
\begin{equation}
C=\frac{-1\pm\sqrt{1+\left( 2\tau_{sf}\gamma \mu E_0 \alpha
\right)^2}}{2\tau_{sf}\gamma \mu E_0 \alpha}A. \label{ccoeff}
\end{equation}
The physical solution corresponds to the + sign in Eq.
(\ref{ccoeff}). It can be easily verified that the solution given by
Eqs. (\ref{s1})-(\ref{s2}), (\ref{ccoeff}) corresponds to $j_y=0$.
Substituting Eq. (\ref{ccoeff}) into Eq. (\ref{eq123}) we finally
get
\begin{equation}
E_y=-\frac{D}{\mu}\alpha-\frac{-1+\sqrt{1+\left( 2\tau_{sf}\gamma
\mu E_0 \alpha \right)^2}}{2\tau_{sf} \mu \alpha}. \label{Ey}
\end{equation}
The second term on the RHS of Eq. (\ref{Ey}) is the electric field
needed to compensate the transverse current arising due to the spin
Hall effect. If we now assume the sample has a finite (but large)
width $L$, then, this term can be interpreted as due to charge
accumulation near the edges, as in the ordinary Hall effect.
Therefore, the Hall voltage can be written as

\begin{eqnarray}
V_H\simeq L \frac{-1+\sqrt{1+\left( 2\tau_{sf}\gamma \mu E_0
\alpha \right)^2}}{2\tau_{sf} \mu \alpha}\approx \nonumber \\
\approx \left\{
\begin{array}{cc}
L\tau_{sf} \mu \alpha\gamma^2E_0^2,&  2\tau_{sf}\gamma \mu E_0
\alpha \ll 1 \\  \\ L\gamma E_0,& 2\tau_{sf}\gamma \mu E_0 \alpha
\gg 1
\end{array}
\right. . \label{VH}
\end{eqnarray}
From this equation we see that the Hall voltage is quadratic in
$E_0$ for small values of the parameter $2\tau_{sf}\gamma \mu E_0
\alpha$, and linear in $E_0$ for large values of this parameter. In
fact, the quadratic dependence is quite unusual, since in the
ordinary Hall effect the Hall voltage is linear in the longitudinal
current. The reason for this unusual dependence can be understood as
follows. The charge current in the $y$ direction, determined by the
difference of spin-up and spin-down currents, has a component
(related to the last term in Eq. (\ref{currentEq})) proportional to
the spin density imbalance $P$ times $\gamma E_0$. At small values
of $2\tau_{sf}\gamma \mu E_0 \alpha$, the spin density imbalance is
proportional to $\gamma E_0$  itself. Therefore, the charge current
and Hall voltage are quadratic in $E_0$. At large values of
$2\tau_{sf}\gamma \mu E_0 \alpha$, the spin density imbalance
saturates and the current dependence on $E_0$ becomes linear.
Another difference with respect to the ordinary Hall effect is that
the polarity of the Hall voltage in the spin Hall effect is fixed by
the geometry of the structure, and does not depend on the direction
of the longitudinal current.

Let us now estimate the magnitude of $2\tau_{sf}\gamma \mu E_0
\alpha$. Taking parameters related to experiments on GaAs
($\tau_{sf}=10$ns, $\gamma=10^{-3}$(Ref.~\onlinecite{Rashba}),
$\mu=8500$cm$^2$/(Vs), $E_0=100$V/cm, $\alpha=2/L$, $L=100\mu$m), we
find $2\tau_{sf}\gamma \mu E_0 \alpha=3.4\cdot10^{-3}$. Therefore,
in experiments with GaAs, most likely, a quadratic Hall voltage
dependence on the longitudinal electric field can be observed.

\begin{figure}[t]
\includegraphics[angle=270,width=8.5cm]{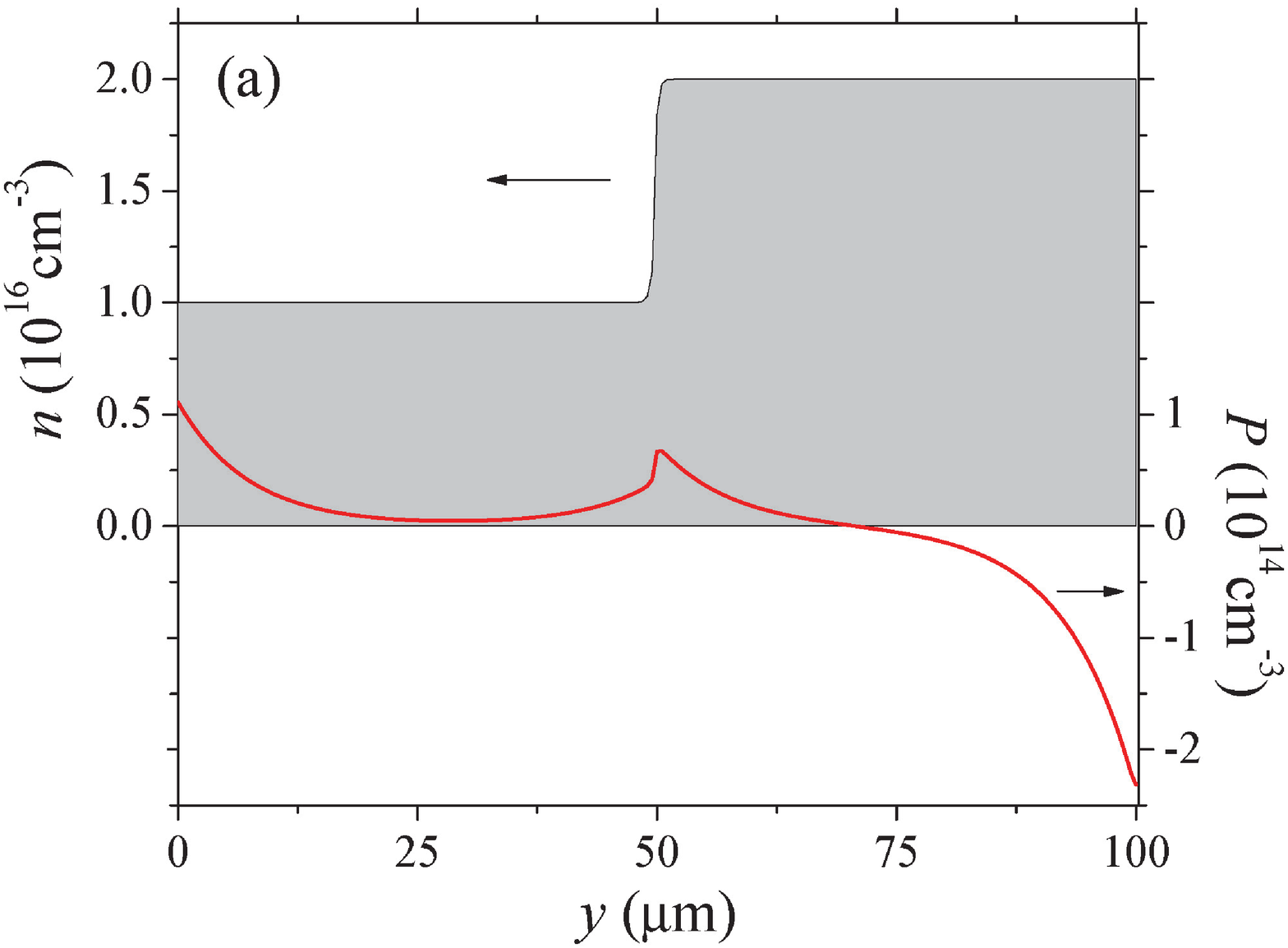}
\includegraphics[angle=270,width=8.5cm]{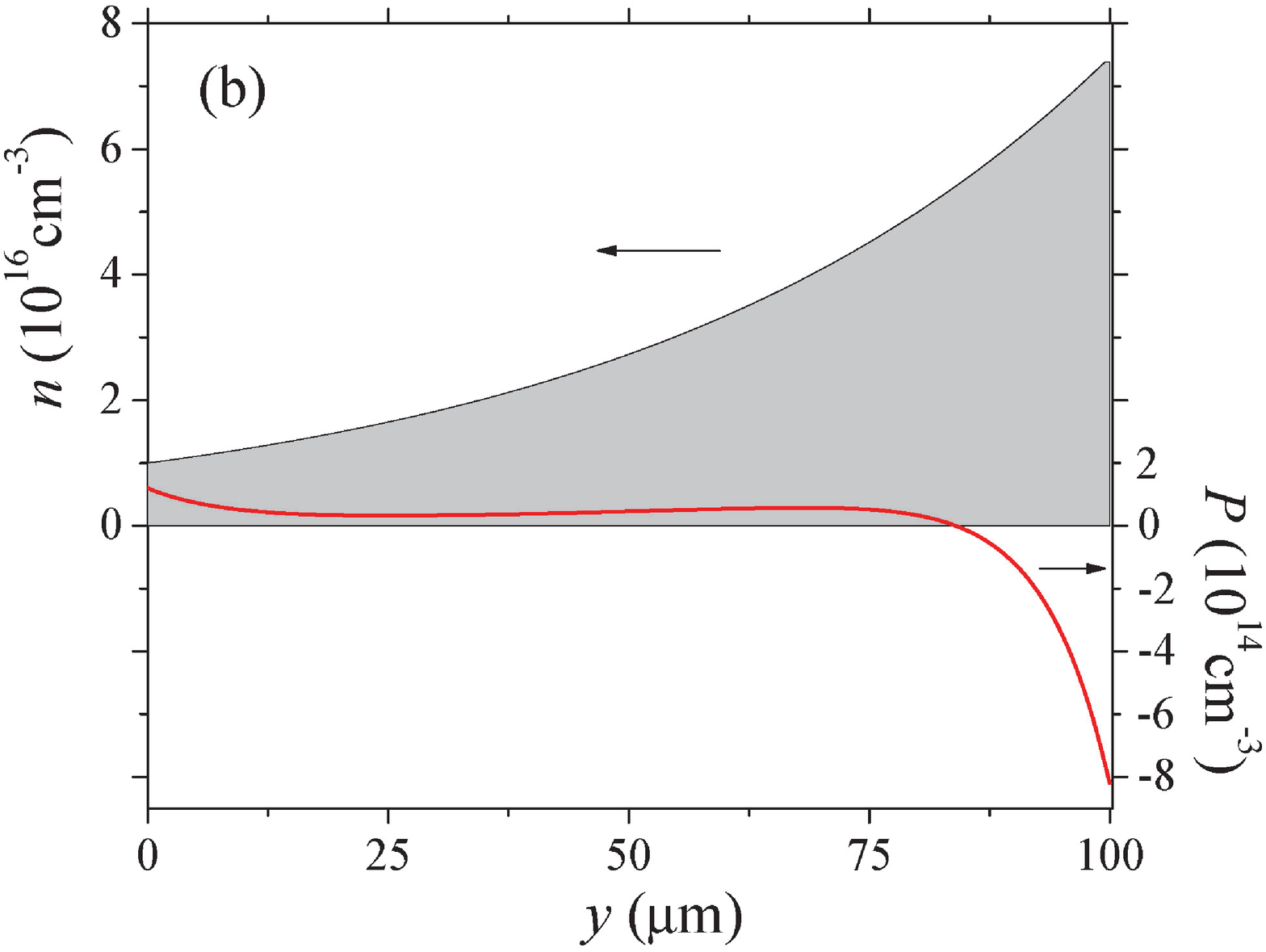}
\caption{\label{fig2}(Color online) Distributions of the electron
density $n(y)$ and spin density imbalance
$P(y)=n_{\uparrow}-n_{\downarrow}$ for a step (a) and exponential
(b) background density profiles. The plots presented in the paper
were obtained using the parameter values
 $\mu=8500$cm$^2$/(Vs), $D=55$cm$^2$/s, $\varepsilon=12.4$,
 $\tau_{sf}=10$ns, $\gamma=10^{-3}$, $E_0=100$V/cm and the background density
 profiles: (a) $N=10^{16}(1+\theta(y-L/2))$cm$^{-3}$ and (b)
 $N=10^{16}\textnormal{exp}(2y/L)$cm$^{-3}$ , where $\theta (..)$ is the
 step function, and $L=100\mu $m is the sample width.}
\end{figure}

{\it Numerical solution --} Equations (\ref{puaeq})-(\ref{Peq}) can
be solved numerically for any reasonable form of $N(y)$. We choose
for their simplicity (and possibility to be realized in practice) a
step profile and an exponential profile. We solve these equations
iteratively, starting with the electron density $n(y)$ close to
$N(y)$ and $P(y)$ close to zero and recalculating $E_y(y)$ at each
time step.~\cite{prec1} Once the steady-state solution is obtained,
the Hall voltage as a function of $E_0$ is calculated as a change of
the electrostatic potential across the sample.

\begin{figure}[t]
\includegraphics[angle=270,width=8.5cm]{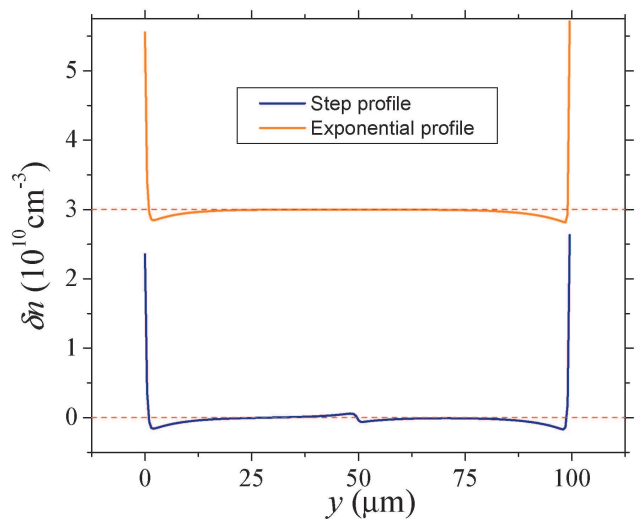}
\caption{\label{fig3}(Color online) Variations in the transverse
charge density induced by the longitudinal current. Here, $\delta
n=n(E_0=100\textnormal{V/cm})-n(E_0=0)$. The curve for the
exponential profile has been shifted vertically by $3\cdot
10^{10}$cm$^{-3}$ for clarity. The dashed lines corresponding to
$\delta n=0$ are there to guide the eye.}
\end{figure}

Fig. \ref{fig2} shows distributions of the charge density and spin
density imbalance in systems with a step (panel (a) of Fig.
\ref{fig2}) and exponential (panel (b) of Fig. \ref{fig2})
background densities. The values of parameters used for these
particular simulations were selected to be close to experimental
conditions reported in Ref. \onlinecite{Kato}. However, we have
tested the robustness of our predictions by solving Eqs.
(\ref{puaeq})-(\ref{Peq}) for different values of parameters, and
found that the predicted Hall voltage should be measurable under a
wide range of experimental parameters. Quite generally, the
self-consistent charge density $n(y)$ is very close to the
background density $N(y)$. Small deviations of $n(y)$ from $N(y)$
can be observed in regions with strong gradients of $N(y)$. In
particular, we can notice that the step profile of electron
density in Fig. \ref{fig2}(a) is smoothed out. Such a charge
redistribution is related to the diffusion term in Eq. (\ref{CC}).
The charge diffusion leads to the formation of a built-in electric
field that equilibrates the charge diffusion.

We also find that the induced spin density imbalance $P$ in systems
with inhomogeneous electron densities shows some new features, in
addition to the well-known spin accumulation near the edges. For
instance, in Fig. \ref{fig2}(a), $P$ has an additional peak around
$y=50\mu$m. In Fig. \ref{fig2}(b), $P$ is almost constant in the
central region of the sample. In both cases, the physics of non-zero
spin density imbalance is the same: the spin-current incoming from
the right is stronger than the spin-current incoming from the left.
We note that the integral spin density imbalance is always zero.

At $E_0=0$, the system is spin-unpolarized and there is no Hall
voltage. When the longitudinal current is switched on, the electron
charge redistributes, and the associated Hall voltage appears. The
change of electron density due to the spin Hall effect is presented
in Fig. \ref{fig3}. The first interesting observation is that there
is a strong charge accumulation near the edges followed by a charge
depletion region. Another observation is that the total electron
density in the left region of the samples ($y<50\mu$m) has increased
and, correspondingly, the total electron density in the right region
has decreased. This change of the electron distribution can be seen
in Fig. \ref{fig3}. Therefore, the left part of the samples is
charged negatively and the right part is charged positively, as
schematically shown in Fig. \ref{fig1}.

The mechanism of formation of sharp peaks of charge accumulation
near the edges is similar to the mechanism of formation of Landauer
resistivity dipoles \cite{Landauer}. From the point of view of spin
currents, the sample edges act as obstacles which block the current
flow, and lead to charge accumulation. The adjacent regions with the
depleted electron density can be interpreted as screening clouds. We
stress that this Landauer-type dipoles of charge accumulation are
quite general for spin Hall systems, and should thus be present also
in traditionally studied structures with a constant density profile.

\begin{figure}[t]
\includegraphics[angle=270,width=8.5cm]{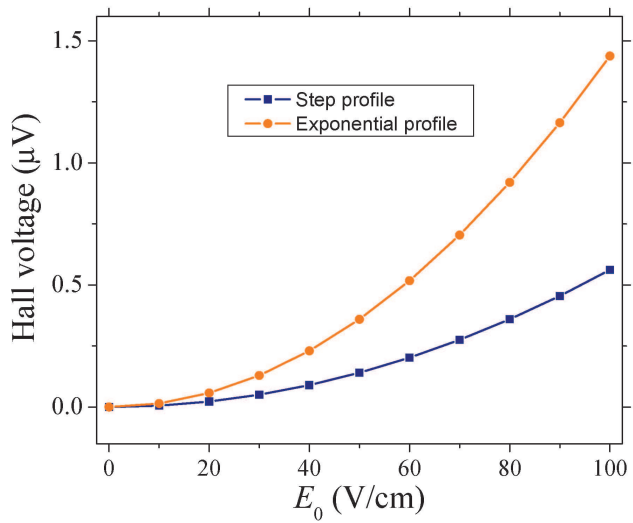}
\caption{\label{fig4}(Color online) Hall voltage as a function of
the longitudinal electric field $E_0$.}
\end{figure}

We finally plot in Fig. \ref{fig4} the change of the electrostatic
potential across the sample as a function of longitudinal electric
field. The Hall voltage, for both density profiles, has a
dependence on $E_0$ which is very close to the exponential
dependence we have predicted analytically in Eq.~\ref{VH} for
small values of $2\tau_{sf}\gamma \mu E_0 \alpha$. The fact that
this exponential dependence appears also in the step profile,
hints at a possible ``general'' property of the Hall voltage in
spin Hall systems with inhomogeneous densities. We emphasize that
a Hall voltage should also appear in spin Hall systems with a
homogeneous electron density, but inhomogeneous $\gamma$. This
corresponds to the case in which the spin-orbit coupling is
dependent on space\cite{serra,nitta}.

In conclusion, we have shown that a Hall voltage would appear in
spin Hall systems with inhomogeneous electron density in the
direction perpendicular to main current flow. The striking result is
that this Hall voltage is generally quadratic in the longitudinal
electric field, unlike the ordinary Hall voltage which is linear in
the same field. These results can be easily verified experimentally,
and would simplify tremendously the measurement of the spin Hall
effect by allowing an electrical measurement of the latter in
non-magnetic systems, and without injection of spin-polarized
electrons.

This work is partly supported by the NSF Grant No. DMR-0133075.


\begin{thebibliography}{}
\bibitem{prec} This is a good approximation for the range of
parameters considered in this work.
\end{thebibliography}
\end{document}